# Cell reorientation under cyclic stretching

Ariel Livne[1], Eran Bouchbinder[2] and Benjamin Geiger[1*]

[1] Department of Molecular Cell Biology and [2] Department of Chemical Physics, Weizmann Institute of Science, Rehovot 7610001, Israel

* Corresponding author. E-mail: benny.geiger@weizmann.ac.il

**Mechanical cues from the extracellular microenvironment play a central role in regulating the structure, function and fate of living cells. Nevertheless, the precise nature of the mechanisms and processes underlying this crucial cellular mechanosensitivity remains a fundamental open problem. Here we provide a novel framework for addressing cellular sensitivity and response to external forces by experimentally and theoretically studying one of its most striking manifestations – cell reorientation to a uniform angle in response to cyclic stretching of the underlying substrate. We first show that existing approaches are incompatible with our extensive measurements of cell reorientation. We then propose a fundamentally new theory that shows that dissipative relaxation of the cell's passively-stored, two-dimensional, elastic energy to its minimum actively drives the reorientation process. Our theory is in excellent quantitative agreement with the complete temporal reorientation dynamics of individual cells, measured over a wide range of experimental conditions, thus elucidating a basic aspect of mechanosensitivity.**

Cells throughout our body constantly interact with their microenvironment. While biochemical communication has been extensively studied for a long time, the importance of mechanical interactions (i.e. cells' ability to apply, sense and respond to forces) has been recognized only recently [1–4]. Precise mechanical conditions, from the subcellular level and up to the organ scale, are critical for tissue development [5,6], function [7,8], remodeling and healing [9,10]. Here we focus on the response of cells to cyclic stretching of the underlying substrate, which mimics vital physiological conditions (e.g. heart beating, pulsating blood vessels and breathing). In response to these external forces, adherent cells

- starting from naturally random orientations - reorient to a well-defined and uniform angle [11] which depends on the applied stretching [12–15]. Moreover, at the subcellular level, the cytoskeleton and most notably stress fibres (SFs) generate internal contractile forces [16] even as they polarize, apparently preceding cell reorientation to a similar angle [14,17]. This outstanding process reveals high cellular sensitivity and accuracy in response to external forces. Nevertheless, the mechanisms underlying it, as well as the validity of current theoretical models describing it [18–22], still remain unclear.

In this study, we experimentally and theoretically study cell reorientation in response to cyclic stretching of the underlying substrate. We first report on detailed experimental measurements of cell reorientation and demonstrate that they cannot be quantitatively explained by the existing models. We then develop a new theory, which takes into account both the passive mechanical response of the cells to substrate deformation and the active remodeling of their actin cytoskeleton and focal adhesions (FAs), highlighting a fascinating interplay between structure, elasticity and molecular kinetics in the reorientation process. This theory is in excellent quantitative agreement with all of the extensive experimental data, predicting the complete temporal reorientation dynamics. Moreover, it elucidates mechanisms involved in cell "readout" of external substrate deformation, an important aspect of cellular mechanosensitivity. Finally, we address the biological and physical significance of the only two cellular parameters appearing in the theory, and discuss the non-trivial predictions that emerge.

## Results

**Reorientation deviates from current theoretical predictions**

We set out first to quantitatively study the reorientation process over a wide range of experimental conditions. REF-52 fibroblasts, which usually grow as polarized cells with long and well separated SFs, were plated onto a fibronectin-coated poly(dimethylsiloxane) (PDMS) chamber. After pre-incubation, the elastic chamber was cyclically stretched, effectively biaxially, in a custom built device [13] at chosen strain amplitudes and defined frequency, f. Specifically, the magnitudes of the linear elastic principal strains in the substrate, $\breve{\varepsilon}_{xx}$ and $\breve{\varepsilon}_{yy}$, were controlled by varying the size, direction and location of the deformation applied at the chamber boundaries.

Cell body reorientation to two mirror-image angles was typically observed within a few hours from the onset of stretching (Fig. 1a-b) (see Supplementary Note 2 for a discussion of the constraints and thresholds for the reorientation process). Similarly, visualization of F-actin, using fluorescent phalloidin, indicated that the SFs of the polarized cells were also aligned along the same angles (Fig. 1c-d). The SF angular distribution following the cyclic stretching (Fig. 1e) had two sharp peaks, corresponding to two mirror-image orientations.

What is the mechanism that determines the final orientation angle, $\bar{\theta}$? A widespread and intuitive approach suggests that the rod-like SFs realign, under cyclic stretching, along the zero (or minimal) matrix strain directions [19–21], where they effectively maintain their original unperturbed state. These zero strain models therefore predict [19]

$$\bar{\theta}_{\text{zero strain}} = \arctan(\sqrt{-\frac{\breve{\varepsilon}_{xx}}{\breve{\varepsilon}_{yy}}}) = \arctan(\sqrt{\frac{1}{r}}) \quad , \tag{1}$$

where we define the biaxiality strain ratio as $r = -\breve{\varepsilon}_{yy}/\breve{\varepsilon}_{xx}$ ($-\infty < r < \infty$). The angle $\theta$ is measured relative to the direction of the principal strain $\breve{\varepsilon}_{xx}$ (in our experiments $\breve{\varepsilon}_{xx}$ is extensional and $\breve{\varepsilon}_{yy}$ compressive with $0 \leq r \leq 1$) (see Fig. 2a). A different approach [18,20], based on measurements of cell traction forces [23], suggests that SFs reorient in the minimal matrix stress direction

$$\bar{\theta}_{\text{zero stress}} = \arctan(\sqrt{-\frac{\breve{\varepsilon}_{xx} + \nu_{\text{substrate}} \cdot \breve{\varepsilon}_{yy}}{\breve{\varepsilon}_{yy} + \nu_{\text{substrate}} \cdot \breve{\varepsilon}_{xx}}}) = \arctan(\sqrt{\frac{1 - \nu_{\text{substrate}} \cdot r}{r - \nu_{\text{substrate}}}}) \quad , \tag{2}$$

where $\nu_{\text{substrate}}$ is the substrate's Poisson's ratio (assuming plane-stress linear elasticity).

To exhaustively test these predictions we performed cyclic stretching experiments with both $\breve{\varepsilon}_{xx}$ and $r$ as our independent control parameters. Consequently, a wide range of final orientations ($\bar{\theta} \approx 45\text{-}80°$) was achieved by modifying the value of $r$ ($r$ was controlled by changing the clamping geometry at the chamber's edges as depicted in Fig. 2b). Surprisingly, the measured angles $\bar{\theta}$ (Fig. 2c) systematically deviate from the zero strain prediction of Eq. 1 (see also [14]), reaching a deviation of ~10 degrees at low $r$ values (20 fold higher than the error bars). An even more dramatic deviation from the zero stress prediction of Eq. 2 is observed (Fig. 2c). Moreover, the statistical variation of the measured final orientations is very narrow (Fig. 1d and Fig. 2c, inset) and cannot account for the discrepancy with the zero strain/stress predictions. We conclude, therefore, that these results call for a new theoretical model.

**New theory of cell reorientation**

The above results demonstrate how SF reorientation depends on the spatiotemporal deformation pattern of the underlying substrate. SFs, however, do not directly interact mechanically with their external environment. Rather, their anchoring to the substrate is mediated via FAs, which are cell adhesion sites that couple to the SF ends. In addition, a network of actin filaments spans between the SFs and interconnects them [24]. These different aspects of SF connectivity and mechanical coupling may be generally modeled by a 2D, anisotropic cellular elastic response. Vis-à-vis the inability of the 1D SF-oriented models to explain our observations and the fact that SF reorientation cannot technically take place without an accompanying change in their adhesion sites, we shift our focus to this broader 2D mechanical system.

Previous theoretical works that extended the scope beyond individual SFs, remained nevertheless confined to 1D geometries. Stamenović *et al.* [22] analyzed the fixed angle solutions for individual, linear SFs with dot-like FAs at their ends. Safran and colleagues, in contrast, addressed cells as 1D force dipoles who actively maintain a stress or strain homeostasis [18,20]. Their model predicted cell reorientation dynamics under cyclic stretching based on a pseudo-energy, not derivable from basic principles, but rather attributed to some biological activity.

We propose that reorientation during cyclic stretching is driven by a dissipative process in which the passively-stored elastic energy of the 2D cell relaxes to a minimum. This smooth rotation takes place through active remodeling and realignment of the relevant molecular structures (which assemble-disassemble in response to forces [23,25]) and continues as long as it is energetically favorable, eventually determining the final orientation angle.

The time-dependent elastic strain energy density stored in the 2D cell (e.g. SF - actin network - FA system or part of it), $u^{\text{cell}}$, is given by [26]

$$u^{\text{cell}} = 0.5 \cdot (\sigma^{\text{cell}}_{\rho\rho}\varepsilon_{\rho\rho} + \sigma^{\text{cell}}_{\theta\theta}\varepsilon_{\theta\theta} + 2\sigma^{\text{cell}}_{\rho\theta}\varepsilon_{\rho\theta}) \ , \qquad (3)$$

where the cell is assumed to inherit the time-varying substrate strains $\varepsilon$ (as the cells are hypothesized to be much softer than the substrate primarily used in this study [22,27,28]) resulting in time-dependent internal stresses, $\sigma^{\text{cell}}$. Note that we address here only the contributions of the time varying strains and stresses due to the cyclic stretch (the much slower internal cell contractility provides an orientation-independent contribution). In addition, we adopt the cell reference frame, with $\rho$ being the direction of cell body, SF and FA polarization [23,29], and $\theta$ was defined above (see Fig. 2a). The total elastic energy stored in a single cell, $U^{\text{cell}}$, is obtained by integrating Eq. 3 over the entire cell area, $A^{\text{cell}}$: $U^{\text{cell}} = \int u^{\text{cell}} \cdot dA^{\text{cell}}$. While $A^{\text{cell}}$, and consequently $U^{\text{cell}}$, may significantly vary between different cells, we assume here that during the reorientation process of individual cells only negligible size changes take place. This implies that the energy minimum depends solely on $u^{\text{cell}}$, making it the relevant physical quantity to analyze. This assumption will be shown below to yield excellent agreement between the theoretical predictions and our experimental measurements.

Substituting the stresses by their anisotropic linear elastic plane-stress expressions, due to the plate-like cell geometry, the stored elastic energy density reads (see Supplementary Note 1)

$$u^{\text{cell}} = 0.5 \cdot K \cdot \varepsilon_{xx}^2 \cdot [\frac{(1+r)\cos^2\theta - 1}{b} + (1-r)]^2 + \text{f}(r) ,\qquad(4)$$

where f(*r*) is a $\theta$-independent function of *r*, and *K* and *b* depend on the cell's anisotropic elastic constants (recall that $u^{\text{cell}}$ and $\varepsilon_{xx}$ are time-dependent). The significance of the dimensionless parameter *b* is better understood through the approximation

$b \approx 1 + \alpha \cdot \frac{E_{\theta\theta}}{E_{\rho\rho}} > 1$, where $E_{\rho\rho}$ and $E_{\theta\theta}$ are the cell's Young's moduli in the ρ and θ directions respectively (see Fig. 2a), and $\alpha$ is of order unity (see Supplementary Note 1 for additional details). In this manner, *b* provides a direct and quantitative measure of the cell's elastic anisotropy. The 1D case, studied by previous theoretical works, is obtained from our 2D theory in the limit of infinite anisotropy ($\frac{E_{\rho\rho}}{E_{\theta\theta}} \to \infty$), resulting in *b*=1.

As the cell's inertia can be neglected, its reorientation is assumed to be driven by simple relaxational dynamics

$$\frac{d\theta}{dt} = -\frac{1}{\eta}\frac{du^{\text{cell}}}{d\theta} ,\qquad(5)$$

where $\eta$ is a viscosity-like coefficient. This suggests that the continuous realignment of cells to a new angle is in fact an ongoing dissipative process of rebuilding and remodeling of the relevant internal structural components that undergo, and consequently drive, reorientation. Furthermore we can replace $\eta$ with $E_{\rho\rho}\cdot\tau$ (by dimensional considerations), where $\tau$ is an intrinsic cell response timescale. This substitution, as will be shown below, highlights $\tau$ as a direct and quantitative measure of cell activity during the reorientation process.

The driving force of the reorientation, according to the right hand side of Eq. 5, is the elastic strain energy pumped into the cells by the cyclic stretching. The cells respond to this force by rotating (left hand side of Eq. 5). This reorientation process takes place through a directed local assembly-disassembly of the relevant cellular molecular components which is controlled by the timescales for recruitment of new molecules or release of bound ones (of the order of 10 s for FAs [30] and the cortical actin network [31]). It is this internal cellular clock, much slower than the external stretch period, which controls the rotation process. Thus, for rapid stretch frequencies ($1/f < 1$s), the driving force may be replaced by its time average (over the characteristic molecular kinetics timescale of ~10s). Using Eq. 4, this yields

$$\frac{d\theta}{dt} = c \cdot \frac{3\breve{\varepsilon}_{xx}^2}{8\tau} \cdot (1+r) \cdot \sin(2\theta) \cdot [\frac{(1+r)\cos^2\theta - 1}{b} + (1-r)] \; , \qquad (6)$$

where $c = \dfrac{K}{b \cdot E_{\rho\rho}}$.

**Direct experimental verification of theoretical predictions**

The major implications of the new theory are encapsulated in Eq. 6, which predicts that reorientation proceeds until reaching one of the two, mirror-image, stable steady state solutions (corresponding to energy minima), $\bar{\theta}_{theory}$

$$\bar{\theta}_{theory} = \arccos(\sqrt{b + \frac{1-2b}{r+1}}) = \arctan(\sqrt{\frac{r + b \cdot (1-r)}{1 - b \cdot (1-r)}}) \ , \tag{7}$$

valid for $1 - 1/b \leq r \leq 1 + 1/(b-1)$.

This prediction is in excellent agreement with the measured $\bar{\theta}$ for a wide range of $r$ values (Fig. 3). We stress that Eq. 7 accurately predicts all of the measured final orientations with a *single* dimensionless parameter, $b = 1.13 \pm 0.04$. Furthermore, experiments performed with very different substrate rigidities (~20kPa in place of the typical ~1MPa), frequencies (1.2-12Hz) or stretch amplitudes (4-24%) all agree with Eq. 7.

The reorientation to the final alignment angle is predicted by this theory to be a continuous rotation process on timescales sufficiently larger than $\tau$. Analyzing the measured, smooth cell orientation dynamics over thousands of stretch cycles (Fig. 4a) we find that Eq. 6 is in excellent agreement with all of our measurements (Fig. 4b) using a *single* timescale $\tau = 6.6 \pm 0.4$s for all cells and experimental conditions analyzed. $\tau$ is independent of initial orientations (Fig. 4b), biaxiality ratios (Fig. 4c), frequencies (1.2-12Hz) and stretch amplitudes (4-24%). The robustness of these results not only lends strong support to our theory, but also indicates that the rotational timescale may be an intrinsic property of cells (possibly cell line dependent).

It is important to understand that while $\tau$ is an intrinsic timescale of the reorientation process, it does not set the overall rotation time, $T$, from the original cell alignment at the onset of cyclic stretching and up to the final orientation angle. As observed in Fig. 4a-c, typical values of $T$ are 1000s, namely 100 times longer than $\tau$. The reason for this marked difference, is that $T$ depends on the product of $\tau$ (~10s), and the externally controlled amplitude $\breve{\varepsilon}_{xx}$ (~0.1 in Fig. 4a-c) thus yielding the typical rotation time: $T \sim \tau/\breve{\varepsilon}_{xx}^2 \sim 10 \cdot 0.1^{-2} \sim 1000$s (see Eq. 6).

Alongside the smooth cell rotation, we encounter a second mode of reorientation. Cells initially co-aligned with the stretching direction often display a loss of polarity soon after the onset of stretch (Fig. 4d). This is quickly followed by a de-novo polarization at an angle close to the final orientation and from there a smooth rotation (described by Eq. 6) to the same alignment as the rest of the cells. This behavior may be related to SF fluidization or rupture under the external stretch [16,32], whereupon the original cytoskeletal elements are also abolished (and consequently cannot drive the rotation).

## Discussion

The theory presented here accurately accounts for the complete reorientation dynamics of individual cells, from an initial random alignment at the onset of cyclic stretching and up to the final orientation angle. It features only two cellular parameters; one related to the cell's 2D anisotropic elastic properties – controlling the final alignment angle – and another to molecular kinetic rates – controlling the reorientation timescale. Both parameters are found constant over the wide range of experimental conditions tested. Namely, the same two values are retrieved when analyzing the smooth rotation of different cells under highly varying stretch conditions. In light of this excellent quantitative agreement, we aim now at exploring the possible biological interpretations of our theory.

SFs, as shown in this work, are clearly involved in cell realignment under cyclic stretching, while disruption of their physical or contractile state, abolishes the reorientation process altogether [17,33]. Nevertheless, the success of our new theory in contrast to previous theoretical works, highlights the role of a 2D, anisotropic elastic response to cyclic stretching, which the 1D SFs cannot account for. This discrepancy could be possibly resolved by taking into consideration the actin filament network which interconnects the different SFs [24]. This not only provides a natural extension to 2D, but also introduces an inherent anisotropic elastic response (SFs being much stiffer than the surrounding actin network [28]) – another important ingredient in our theoretical picture. In addition, FAs may also play a key role in the reorientation process as they anchor and couple the SFs to the cyclically deforming external environment. In light of their well-established mechanosensitivity [4], 2D geometry and anisotropic structure [34,35] it is plausible that the FAs themselves reorient under cyclic stretching, driving the observed SF rotation in their wake (note that SFs, cannot rotate per se, without an accompanying change in their adhesion structures). Next, we discuss a possible interpretation of our measurements in terms of a cell response driven by a combined SF - actin network - FA system, or part of it (FAs alone or SFs - actin network alone).

The two parameters extracted by our theory provide a quantitative insight into the structure and dynamics of the cell component(s) driving the realignment process. The constancy of the reorientation timescale, $\tau$, (6.6 ± 0.4s) under widely varying stretch conditions, indicates that this parameter is an intrinsic cellular property, possibly also related to the timescale of FA-transmitted traction fluctuations [36]. At the FA level, rotation is a coordinated process of assembly-disassembly, absorbing protein components from the immediate cytoplasm at one end and disintegrating at the other end. It is limited, therefore, by the molecular kinetics of the constituent building blocks. Similarly, effective SF rotation may take place by de-novo assembly and remodeling through the underlying actin filament network [37,38]. Direct measurements of such FA and cortical actin molecular kinetic times from photobleaching experiments [30,31] closely match the $\tau$ value extracted from rotation measurements. Thus, $\tau$ may serve as an indirect measure of molecular reaction timescales for proteins responding to force perturbations. Identifying the precise protein(s) responsible for the timescale $\tau$ is a particularly appealing direction for future investigation in light of the immense complexity of the FA molecular structure, which spans over a hundred different proteins and almost a thousand interactions [39]. Interestingly, knockout of vinculin, a key FA protein, results in embryonic death with heart orientation defects [40].

The parameter $b$, related to the cell's (or relevant part of it) elastic moduli (see Supplementary Note 1), attains a value (1.13 ± 0.04) that suggests significant - yet finite - elastic anisotropy, with about ten fold stiffer mechanical response along its long axis compared to the lateral axis. This is in agreement with AFM measurements showing that SFs are almost ten fold stiffer than the surrounding cytosol [28]. In addition, while no direct measurements of FA elastic properties are available, high resolution structural studies of FAs are also consistent with this result, revealing significant differences between the adhesions' long and short axes [34,35]. Finally, the independent experiments of [14], performed on different cells, are in excellent agreement with our theory and *b* parameter (Fig. 3). This may suggest that the elastic properties, mirrored by *b*, are cell line and substrate independent, indicating a potential universality in cell elastic response. One possible mechanism that can account for a widespread formation of cell structures with similar elastic properties is through equilibrium self-assembly (as suggested e.g. in [41]).

The quantitative agreement between our extensive measurements and new theory elucidates an important aspect of cellular mechanosensitivity. Over the last two decades, considerable progress has been achieved towards understanding the "inside-out" role of FAs and SFs in force transmission to the extra cellular matrix [23,25,42,43]. In contrast, the reverse mechanisms involved in cell "readout" of external substrate deformations, through these same structures, still remain unclear. To this effect, the cyclic stretching experiments described here, shed light on the role of elastic energy minimization in driving an "outside-in" cell-level response of reorientation and directed migration [17]. In addition, the success of the dissipative dynamics approach adopted here may offer insights into the physiological motivation for the reorientation process. One possible rationale is that cell alignment at an orientation of minimal elastic energy provides an optimal configuration to minimize energy expenditure by the body. In this manner cells can reside in regions of high deformation and contractility while posing a minimum mechanical load on the elements driving the cyclic stretching itself (e.g. heart, lungs).

In conclusion, we provide here a novel framework for addressing and understanding cellular mechanosensitivity. A new, biologically and physically motivated, mathematical description of the mechanism and dynamics underlying cell reorientation under cyclic stretching, which is regulated and driven by the mechanoresponsive SF-FA system, is developed. This theory is strictly based on the molecular and physical properties of SFs and FAs, and hence constitutes a new first-principles approach which significantly enhances our understanding of cellular mechanosensing. Moreover, it offers quantitative tools with predictive powers that are relevant to a variety of cell behaviors with potential applications in tissue engineering and biomedicine.

## Methods

**Cell culture**

REF-52 cells stably expressing YFP-tagged paxillin, previously described in [44], were grown in Dulbecco's modified Eagle's medium supplemented with 10% fetal bovine serum, 2mM glutamine, 100 U/ml penicillin and 100μg/ml streptomycin. The same medium was also used for time-lapse microscopy in a 5% $CO_2$ humidified atmosphere at 37°C. All cell culture components were provided by Biological Industries, Beit Haemek, Israel.

**Cell stretching**

A custom built stretching device, developed by Martin Deibler and Ralf Kemkemer from the MPI, Stuttgart, Germany, was used to cyclically strain an elastomeric membrane that served as cell culture substrate [13]. The membrane was stretched back and forth by a brushless servo motor (Faulhaber) with an attached 14:1 gear unit, through a setup of eccentric tappet and conrod. Stretch amplitude (0.1-30%) and frequency (0.001-15Hz) were controlled by the choice of eccentric used and of motor rotation speed. The mechanical stretcher was mounted in an upright microscope (Zeiss Axiophot equipped with Zeiss x10/0.3W and x40/0.8W objectives) on a specially adapted mechanical XY stage. In this manner different cell regions on the stretched membrane could be analyzed. Image-Pro Plus 7.0 software (Media Cybernetics) controlled image acquisition by a CCD camera (PCO Pixelfly) with 1392x1040 pixel resolution (6.45μm pixel size), the illumination shutters (Uniblitz), Z axis focusing motor (Marzhauser) and the stretcher motor. The microscope was placed in a custom built chamber under controlled temperature and $CO_2$ concentration. Time lapse images of the cell reorientation were acquired every 2 minutes, during short arrest of the stretch cycles, at regions of uniform *r* and using the x10 objective.

**PDMS substrate**

Cells were plated on a 20µg/ml fibronectin-coated (Sigma-Aldrich) PDMS chamber (Sylgard 184, Dow Corning). Typically, ~15,000 cells were seeded on the 2cm x 2cm membrane, at the bottom of the chamber, 16-24h before stretch application. The substrate's Young's modulus is estimated at ~1MPa [13] for the 10:1 (base: curing agent) PDMS primarily used in this study. To obtain a lower rigidity (~20kPa), a second, ~100µm thick layer of 50:1 PDMS [23] was spin coated (WS-650MZ-23NPP/Lite, Laurell Technologies) on top of a plasma treated (Harrick Plasma) 10:1 chamber. The plasma treatment was used to prevent diffusion of the soft layer into the underlying substrate during the curing process.

**Displacement field measurements and strain field calculations**

The displacement fields in the PDMS substrate due to the cyclic stretching were directly measured in the following manner. Two snapshots of the cells in the chamber were acquired by phase microscopy: one immediately preceding the stretch and another at the maximal stretch. A custom particle-tracking code written in Matlab (MathWorks) paired the cells in the two images by cross correlating small boxes (typically 40 pixels in length at 20 pixel intervals). The peak correlation for each box gave a sub-pixel accuracy for the displacement field generated by the substrate stretch. Finally, differentiation of the displacement field yielded the strain field [26]. This was performed for each experiment, both before the onset of cyclic stretching and at the end, prior to fixing the cells.

For comparison, the displacement field was also measured using submicron fluorescent tracer particles embedded in the PDMS directly below the cells, as well as using scratches on the substrate surface itself. These measurements provided practically identical results to the cell correlation method.

**SF analysis**

At the end of the cyclic stretching experiments, once reorientation was complete, cells were fixed for two minutes in 37°C warm 3% paraformaldehyde (PFA) in phosphate-buffered saline (PBS) containing 0.5% Triton X-100, and then post-fixed with PFA alone for an additional 20 minutes. The cells were then washed three times with PBS and stained with TRITC-labeled phalloidin (Sigma-Aldrich). Next, we returned at higher magnification (x40 compared to x10), to the region of interest that was tracked during the experiment and acquired fluorescence images of the SFs and FAs of the individual cells. At the final step, the individual SFs were segmented and their orientation analyzed by a custom Matlab code implementing Zemel et al.'s algorithm [45].

**Measuring the final orientation angle**

The final orientation angle, $\bar{\theta}$, was measured at the end of the cyclic stretching experiments, once reorientation was complete, in one of two ways. In the region of interest, where *r* was uniform, the mean of the different cell body orientations (n>30) was taken. Post-mitotic cells as well as cells in contact with one another were discarded from this analysis. Alternatively, the peak of the SF angular distribution of these cells was used (e.g. Fig. 1e). Comparing the results of these two methods, we found them practically identical, with an advantage for the SF analysis due to its smaller measurement uncertainty. We conclude, therefore, that both SFs and cell bodies reorient under cyclic stretch, to the same, well defined angle (Fig. 1d) and that both measurement techniques could be used interchangeably.

## Acknowledgements

We thank M. Deibler and R. Kemkemer for design, setup and guidance of the cell stretcher machine. We would also like to thank Z. Kam for numerous discussions and careful review of this manuscript and S. Safran and A. Bershadsky for useful comments. This research was supported by funding from the European Union's Seventh Framework Programme, European Research Council (ERC) Advanced program, under grant agreement n$^o$ 294852 (SynAd). E.B. acknowledges support from the Harold Perlman Family Foundation and the William Z. and Eda Bess Novick Young Scientist Fund. B.G. is the incumbent of the Erwin Neter Professorial Chair in Cell and Tumor Biology.

## Author contributions

A.L. and B.G. designed the experiments. A.L. performed the experiments and analysed the data. A.L. and E.B. developed the theory. All authors contributed to the writing of the manuscript.

**Competing financial interests:** The authors declare no competing financial interests.

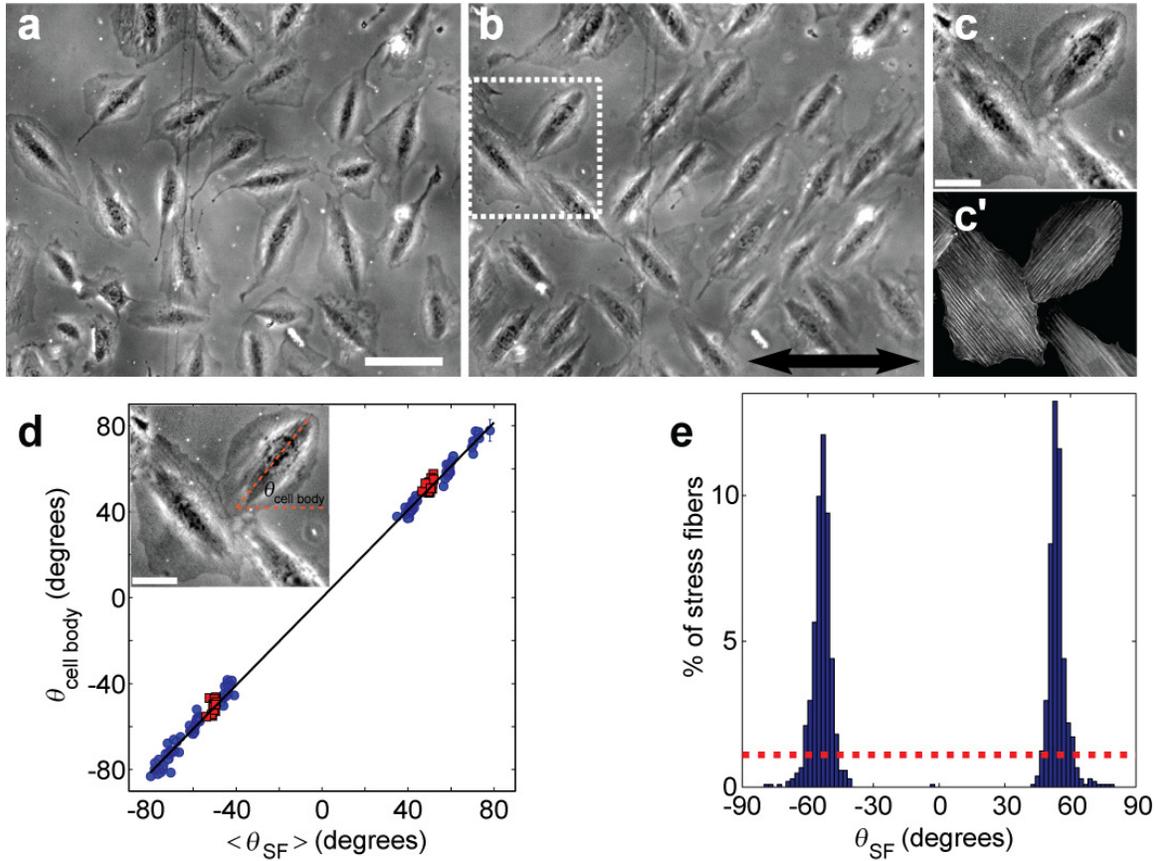

**Figure 1 - Cyclic stretching reorients cells and SFs along two mirror-image angles.**

**a-b**, Phase-contrast images of REF-52 fibroblast cells on a fibronectin-coated PDMS substrate, before (**a**) and after (**b**) 6 hours of cyclic stretching (10% strain at 1.2Hz), show reorientation from random cell alignments to two, well defined, mirror-image angles. The largest principal strain (stretch) was applied in the horizontal direction as shown by double sided arrow. Bar = 100μm. **c-c'**, Closeup of reoriented cells (**c**) shows SF alignment (**c'**) at a similar angle to the cell body. SFs were imaged after being stained with fluorescently labeled phalloidin. Bar = 40μm. **d**, The mean SF orientation of individual, polarized, cells ($<\theta_{SF}>$) matches the cell body orientation ($\theta_{cell\,body}$). Data from different experiments (red squares correspond to cells from (b)) yield a linear relation of slope ~1 between the two angles (black line is the best fit: $\theta_{cell\ body} \approx 1.02 \cdot \theta_{SF}$). The final orientation angle varies due to the cyclic stretching conditions (see Fig. 2 for more details). Inset shows how $\theta_{cell\ body}$ was determined in phase-contrast images as the long axis angle of the dark, actin-rich, cell core. Error bars represent 95% confidence intervals. Bar = 40μm. **e**, Analysis of individual SF orientations, $\theta_{SF}$, at the end of the cyclic stretching (~1000 SFs from (b) and its vicinity) reveals an angular distribution with two sharp peaks, as contrasted with initial random configuration (dashed red line).

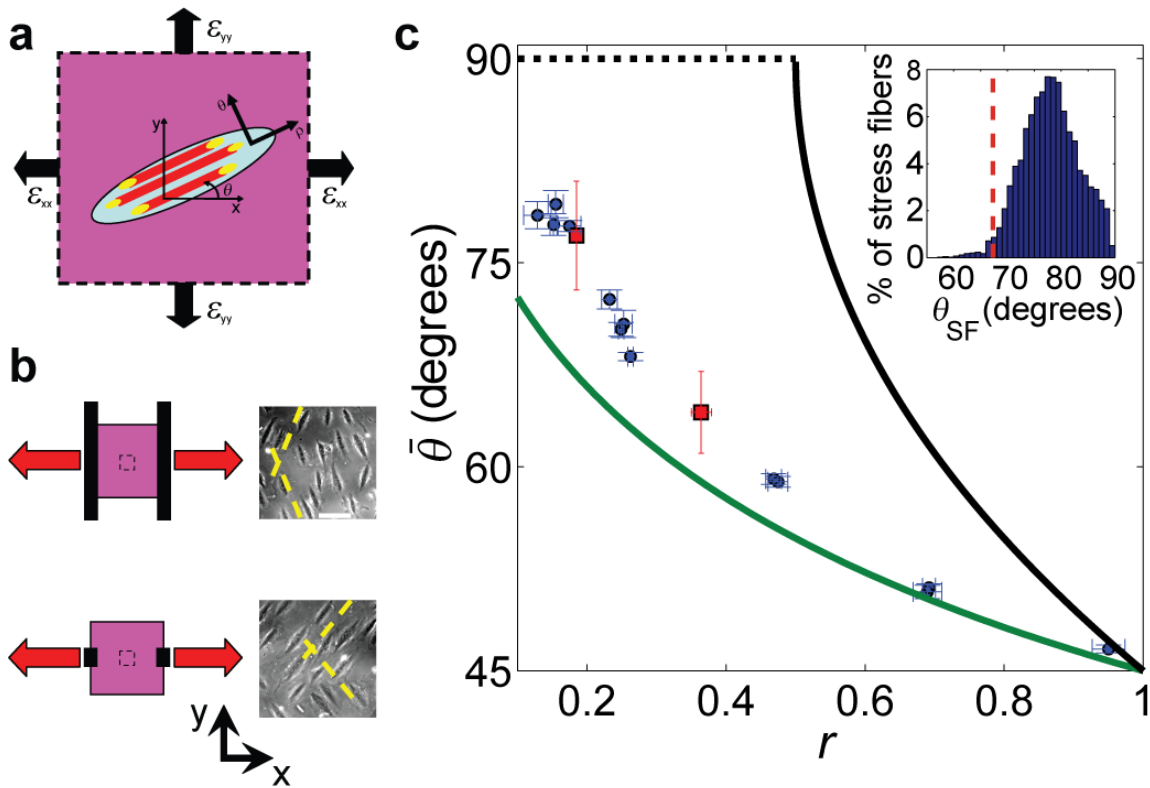

**Figure 2 - The final orientation angle is determined by the strains in the underlying substrate and differs from previous theoretical models.**

**a**, Cartoon of a single cell (light blue ellipse) on a deformable 2D substrate (magenta), that is stretched with principal strains: $\varepsilon_{xx}$ & $\varepsilon_{yy}$ (in our experiments $\varepsilon_{xx}$ is extensional and $\varepsilon_{yy}$ compressive). ρ marks the direction of cell body, SF (red) and FA (yellow) polarization which is at angle, $\theta$, relative to the direction of the principal strain $\varepsilon_{xx}$. **b**, Schematic presentation of different loading and clamping conditions (left) and phase-contrast images of typical cell orientations at the end of the stretch cycles (right). The 2cm x 2cm PDMS substrate depicted in magenta, is stretched (red arrows) via clamps (black solid lines) attached at its boundaries. By adjusting the clamps' location and size we could tune the strains transferred to the ~1 mm$^2$ region of interest (inner dashed box), at the substrate's center. In this manner we could control the final cell orientation (dashed yellow lines are guides to the eye). Bar = 100μm. **c**, The measured final orientation angle, $\bar{\theta}$, as a function of the biaxiality ratio, *r*. Each point (blue circles = SF orientations, red squares = cell body orientations) was extracted from a different experiment (1.2Hz, 4-24% strain) and represents the mean angle for the relevant cell population (n>30) in the region of interest. The green and black lines are respectively the zero strain (Eq. 1) and zero stress (Eq. 2) theoretical predictions. The dashed black line is the minimal stress prediction which extends the zero stress prediction to regions where the latter has no solution. Error bars represent 95% confidence intervals. Inset shows the SF angular distribution of a single experiment. Note that the zero strain prediction (dashed red line) is an outlier in the measured distribution, and cannot account for the discrepancy observed.

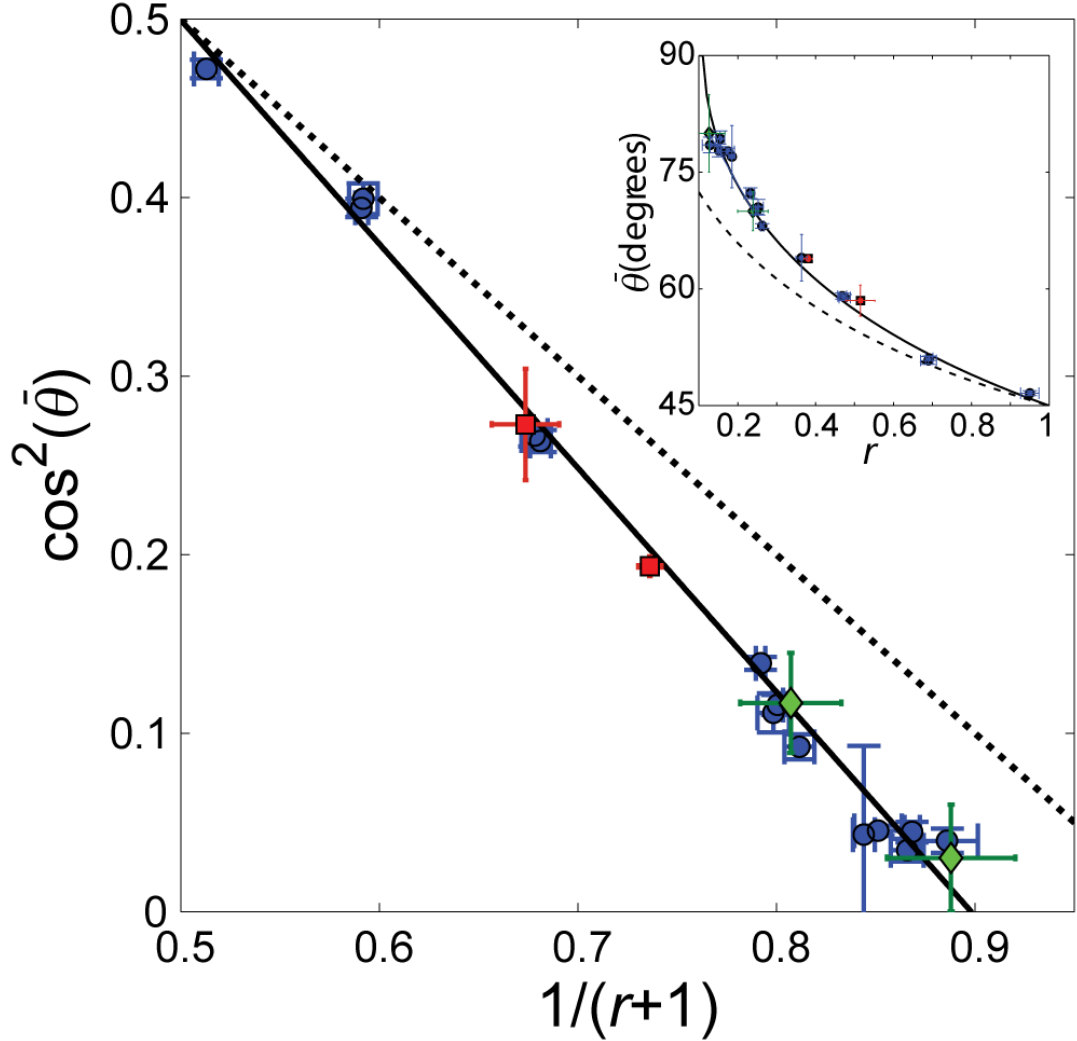

**Figure 3 - Final orientation angle is correctly captured by the proposed theory.**

As predicted by the theory (Eq. 7), a linear relation between $\cos^2(\bar{\theta})$ and $(r+1)^{-1}$ is clearly observed, where $\bar{\theta}$ is the measured final orientation angle (blue circles are data from Fig. 2c). This excellent agreement (solid black line is the best fit to Eq. 7) depends on a *single* parameter, $b$, where both the slope and the intercept are uniquely determined by it ($b = 1.13 \pm 0.04$ extracted from fit). In comparison, the zero strain prediction (Eq. 1) is depicted by the dashed black line. Additional measurements - performed on a much softer substrate (~20 kPa compared to ~1MPa) (red squares) as well as data extracted from the literature for a different cell line [14] (green diamonds) – fall on the same line. This suggests that the elastic properties, associated with the b parameter in our theory, do not depend on substrate stiffness and are possibly cell line independent. Inset shows the same data, best fit and zero strain prediction, as above, with $\bar{\theta}$ plotted directly vs. *r*. Error bars represent 95% confidence intervals.

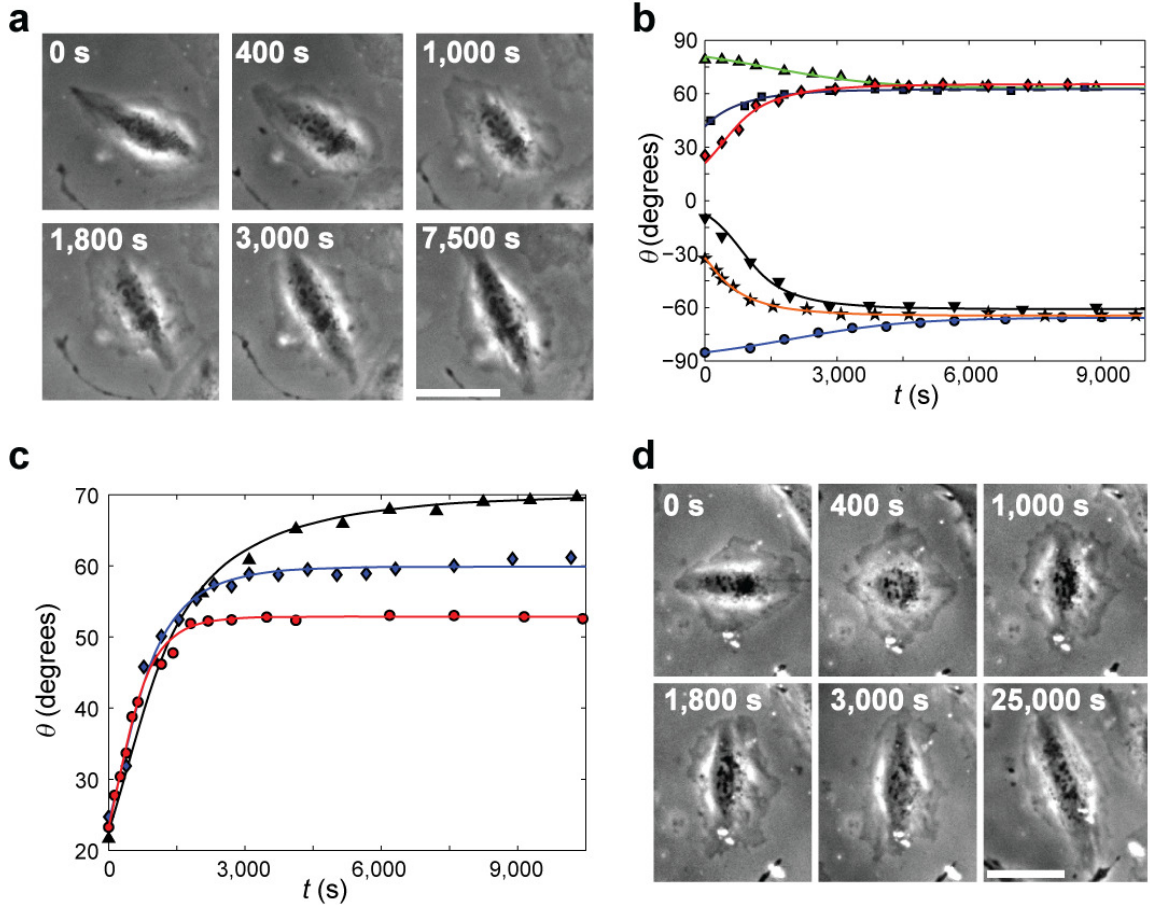

**Figure 4 - Reorientation dynamics are quantitatively explained by the proposed theory.**

**a**, Phase contrast snapshots tracking a single cell reorientation dynamics under cyclic stretching ($r=0.38$, $\breve{\varepsilon}_{xx}=0.11$ at 1.2Hz). The elapsed time from the beginning of cyclic stretching is marked on each image. **b**, Cell body orientations, $\theta$, of six cells, originally polarized in different directions, were recorded from the onset of cyclic stretching, $t=0$ ($r=0.36$, $\breve{\varepsilon}_{xx}=0.11$ at 1.2Hz). Reorientation takes place by a smooth rotation towards the closer of the two mirror-image final alignment angles (here: $\pm 64°$). The individual dynamics leading up to these set points strongly depends on the initial orientation. The reorientation duration is not a simple function of the total rotation angle. Comparing the recorded reorientations to the theory's predictions (Eq. 6) we find that individual best fits (solid curves) are not only in excellent agreement with measurements, but also all yield the same $\tau=6.6\pm0.4$s value ($b=1.13$, independently extracted from Fig. 3, and $c=1$, as explained in the Supplementary Note 1, were used in the analysis). **c,** Cells initially oriented at a similar initial angle rotate towards different final orientations according to the applied biaxiality ratio (triangles: $r = 0.25$, diamonds: $r = 0.48$, circles: $r = 0.69$; $\breve{\varepsilon}_{xx} \approx$ 0.10 at 1.2Hz for all three). The theory accurately describes the reorientation dynamics and predicts the same $\tau \sim 6.6$s value for different cells under a wide range of experimental conditions (solid curves are single parameter fits to Eq. 6). Therefore, analysis of the smooth reorientation of a single cell towards the final orientation predicts the rotational dynamics of all other cells, even when stretched under widely different experimental conditions. **d,** Phase contrast snapshots tracking a single cell initially co-aligned with the stretching direction. The cell loses polarity shortly after the onset of cyclic stretching ($t=400$s). This is quickly followed ($t=1000$s) by a de-novo polarization at an angle close to the final orientation from which the cell smoothly rotates to $\bar{\theta}$ (stretch parameters as in (a)). In **a** & **d**, the largest principal strain (stretch) was applied in the horizontal direction and bar = 50μm.